# A FINITE ELEMENT MODEL AND ELECTRONIC ANALOGUE OF PIPELINE PRESSURE TRANSIENTS WITH FREQUENCY-DEPENDENT FRICTION

**J.-J. Shu**[1]

## ABSTRACT

A finite element model and its equivalent electronic analogue circuit of hydraulic transmission lines have been developed. Basic equations are approximated to be a set of ordinary differential equations that can be represented in state space form. The accuracy of the model is demonstrated by comparison with the method of characteristics.

## INTRODUCTION

If external perturbations are superimposed onto a steady system, the equilibrium no longer exists. Those perturbations will propagate in the form of waves. Some of waves chase each other to form weakly-discontinuous rarefaction waves and strongly-discontinuous shock waves. In hydraulic systems, flow transients typically occur when there is either a retardation of the flow due to the closure of a pump valve or an acceleration due to the opening of the valve. This may cause damage to hydraulic components, reduce volumetric efficiency and hence disturb normal operations by forming the rarefaction wave, commonly known as waterhammer, and the shock wave, generated by cavitation.

Dynamic analysis in a time domain is one of the most important parts in computer simulation for hydraulic systems. Most transient distributed models can be described by hyperbolic partial differential equations. The updated-accurate numerical scheme is the method of characteristics (Gray, 1953; Streeter and Lai, 1962; Evangelisti, 1965, 1969) which has widely and successfully been used in tackling the problems of fluid transients such as waterhammer (Wylie and Streeter, 1978; Shu et al. 1997) and cavitation (Shu, 2003). In the calculation process of the method of characteristics, due to the determination of the time step and space step, there are certain difficulties in determining the time step and boundary conditions compatible with other hydraulic lumped or distributed models, especially in the designs of expert systems, large-scale power system packaging and parallel processing. For example, at least $2000$ computing elements are needed for a time step with $10^{-2}$ millisecond along a tube $20\,\mathrm{m}$ in length. When variable time steps are required, the calculation for intermediate interpolations becomes very expansive. Another disadvantage of using the method of characteristics is that some mathematical transformations are required before a numerical scheme is implemented.

In short, the purpose of this paper is to illustrate how the Galerkin finite element method (Galerkin, 1915) can be applied only to space variables by taking a simple and well-defined problem as an example, and to give rise to an initial value problem for a system of ordinary differential equations in order to decouple time step from space step. Meanwhile an equivalent electronic analogue circuit is established. There is no difficulty in following the presented procedures to solve other waterhammer problems and even cavitation problems.

## BASIC GOVERNING EQUATIONS

[1] School of Mechanical & Aerospace Engineering, Nanyang Technological University, 50 Nanyang Avenue, Singapore 639798

Before introducing the procedure itself, I would first like to state a physical problem to which it will be applied. Because of sudden valve closure at the start of a fluid transmission line, the flow within the transmission line is to be calculated under the assumptions of one-dimensional, unsteady and compressible flow. The independent variables of space and time are denoted $x$ and $t$. The dependent variables are $P$, the pressure and $Q$, the flowrate. In the analysis of fluid transients, two basic principles of mechanics namely (a) the principle of the law of conservation of mass and (b) the principle of the law of Newton' momentum give rise to two partial differential equations.

(a) Equation of Continuity

$$\frac{1}{c_0^2}\frac{\partial P}{\partial t}+\frac{\rho}{\pi r_0^2}\frac{\partial Q}{\partial x}=0 \tag{1}$$

(b) Equation of Motion

$$\frac{\rho}{\pi r_0^2}\frac{\partial Q}{\partial t}+\frac{\partial P}{\partial x}+F(Q)+\rho g\sin\theta_0=0 \tag{2}$$

where $c_0$ is the acoustic velocity, $\rho$ is the density, $\mu$ is the viscosity, $r_0$ is the internal radius of the transmission line tube, $\theta_0$ is the angle of the transmission line tube inclined with the horizontal. The friction term $F(Q)$ can be expressed as a steady friction term $F_0$ plus a frequency-dependent friction term, for which a model has been developed by Zielke (1968) and Kagawa et al. (1983).

$$F=F_0+\frac{1}{2}\sum_{i=1}^{k}Y_i \tag{3}$$

$$\begin{cases}\dfrac{\partial Y_i}{\partial t}=-\dfrac{n_i\mu}{\rho r_0^2}Y_i+m_i\dfrac{\partial F_0}{\partial t}\\ \qquad Y_i(0)=0.\end{cases} \tag{4}$$

The constants $n_i$ and $m_i$ are tabulated in Table 1 and the number of terms $k$ to be selected should be determined according to the relation between the break frequency of the approximated weighting function and the frequency range of the system, which is assumed in steady-state when time $t=0$.

Table 1: $n_i$ and $m_i$

| $i$ | 1 | 2 | 3 | 4 | 5 | 6 | 7 | 8 | 9 | 10 |
|---|---|---|---|---|---|---|---|---|---|---|
| $n_i$ | $2.63744\times10^1$ | $7.28033\times10^1$ | $1.87424\times10^2$ | $5.36626\times10^2$ | $1.57060\times10^3$ | $4.61813\times10^3$ | $1.36011\times10^4$ | $4.00825\times10^4$ | $1.18153\times10^5$ | $3.48316\times10^5$ |
| $m_i$ | 1.0 | 1.16725 | 2.20064 | 3.92861 | 6.78788 | $1.16761\times10^1$ | $2.00612\times10^1$ | $3.44541\times10^1$ | $5.91642\times10^1$ | $1.01590\times10^2$ |

## GALERKIN FINITE ELEMENT METHOD

The finite element formulations with use of the Galerkin method in time domain analysis were presented by Rachford and Ramsey (1975) and Paygude et al. (1985) using a conventional uniformly spaced grid system with two degrees of freedom, pressure and flowrate; however, this method produced many numerical oscillations due to the lack of frequency-dependent friction. In the work that follows, the Galerkin finite element method to fluid transients is re-examined using one degree of freedom, pressure or flowrate.

Let $U=\dfrac{\rho}{\pi r_0^2}Q$, $F_0(Q)=\dfrac{8\mu}{\pi r_0^4}Q$, $R=\dfrac{8\mu}{\rho r_0^2}$ and $H_0=\rho g\sin\theta_0$ for the case of linear steady friction, the equations (1), (2) and (4) can be rearranged according to the operator equations

$$L_1(U,P,Y_i) \equiv \frac{1}{c_0^2}\frac{\partial P}{\partial t} + \frac{\partial U}{\partial x} = 0$$

$$L_2(U,P,Y_i) \equiv \frac{\partial U}{\partial t} + \frac{\partial P}{\partial x} + RU + \frac{1}{2}\sum_{i=1}^{k} Y_i + H_0 = 0$$

$$L_3(U,P,Y_i) \equiv \frac{\partial Y_i}{\partial t} + \frac{n_i R}{8} Y_i - m_i R \frac{\partial U}{\partial t} = 0.$$

When the transmission line is divided into $2N+1$ equal elements, each $\Delta x$ in length, as shown in Figure 1 and a suitable finite dimensional space is spanned by shape functions (Galerkin, 1915)

$$w_j^+(x) = \frac{x_{j+2} - x}{x_{j+2} - x_j} \qquad w_j^-(x) = \frac{x - x_j}{x_{j+2} - x_j}.$$

The Galerkin method consists in finding approximations to $U$, $P$, and $Y_i$ of the form

$$U(x,t) = u_{2j}(t) w_{2j}^+(x) + u_{2j+2}(t) w_{2j+2}^-(x)$$

$$P(x,t) = p_{2j+1}(t) w_{2j+1}^+(x) + p_{2j+3}(t) w_{2j+3}^-(x)$$

$$Y_i(x,t) = y_{i,2j}(t) w_{2j}^+(x) + y_{i,2j+2}(t) w_{2j+2}^-(x)$$

for $j = 0,1,\cdots,N-1$, where the unknown coefficients $u$, $p$, and $y_i$ are the nodal values of $U$, $P$, and $Y_i$, respectively, determined by the inner product $(\bullet,\bullet)$, so

$$\left(L_i, w_j^+\right) \equiv \int_{x_j}^{x_{j+2}} w_j^+ L_i dx = 0 \quad \text{and} \quad \left(L_i, w_j^-\right) \equiv \int_{x_j}^{x_{j-2}} w_j^- L_i dx = 0$$

when $i = 1$, $j$ is selected as an odd number, and $i = 2$ or $3$, $j$ is selected as an even number.

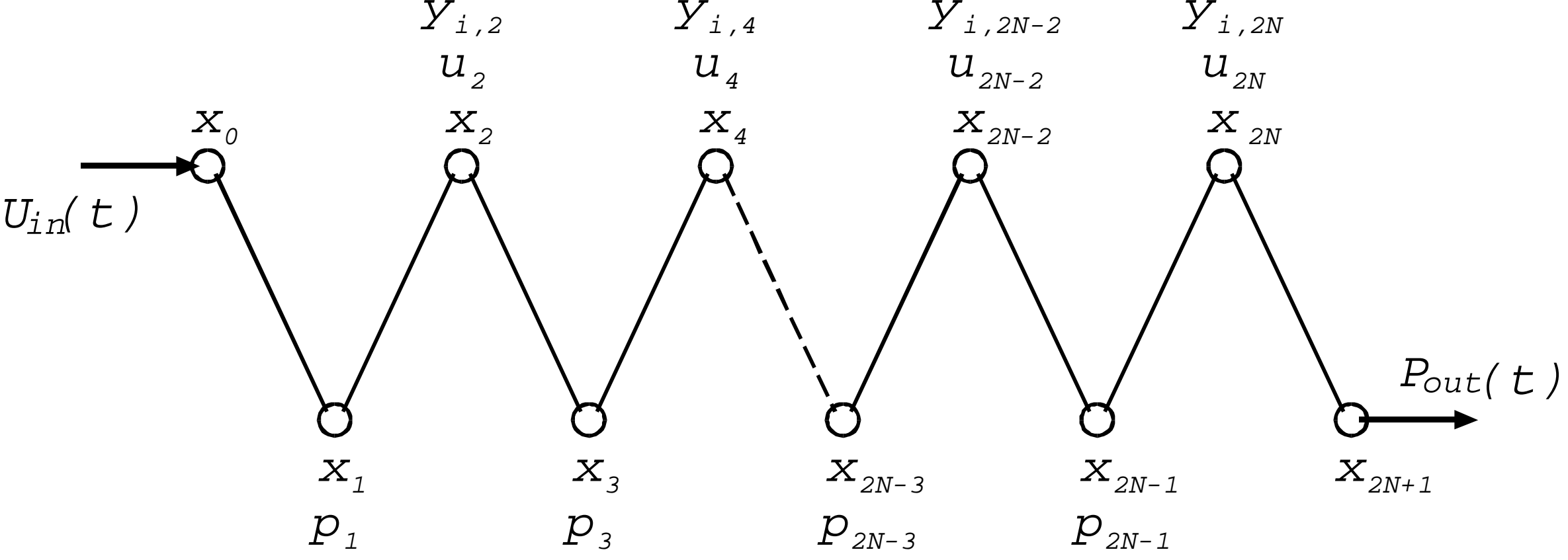

Figure 1: Interlacing grid system

Considering the boundary conditions where the flowrate $u_0 = U_{\text{in}}(t)$ at one end is known and the pressure $p_{N+1} = P_{\text{out}}(t)$ at another end is known, the resultant ordinary differential equation is of the following form

$$
\frac{d}{dt}\begin{pmatrix}\tilde{\mathbf{u}}\\ \tilde{\mathbf{p}}\\ \tilde{\mathbf{y}}\end{pmatrix}=\begin{pmatrix}-R\mathbf{I} & \dfrac{1}{16\Delta x}\mathbf{A} & -\dfrac{1}{2}\left(\tilde{\boldsymbol{\Xi}}\otimes\mathbf{I}\right)^{T}\\ \dfrac{c_0^2}{16\Delta x}\mathbf{B} & \mathbf{O} & \mathbf{O}\\ -R^2\tilde{\mathbf{M}}\otimes\mathbf{I} & \dfrac{R}{16\Delta x}\tilde{\mathbf{M}}\otimes\mathbf{A} & -\dfrac{R}{8}\left[\mathbf{F}+4\tilde{\mathbf{M}}\otimes\left(\tilde{\boldsymbol{\Xi}}\otimes\mathbf{I}\right)^{T}\right]\end{pmatrix}\begin{pmatrix}\tilde{\mathbf{u}}\\ \tilde{\mathbf{p}}\\ \tilde{\mathbf{y}}\end{pmatrix}+\frac{c_0^2}{16\Delta x}U_{\mathrm{in}}\begin{pmatrix}\tilde{\mathbf{O}}\\ \tilde{\mathbf{C}}\\ \tilde{\mathbf{O}}\end{pmatrix}
$$

$$
+\frac{1}{16\Delta x}P_{\mathrm{out}}\begin{pmatrix}\tilde{\mathbf{D}}\\ \tilde{\mathbf{O}}\\ R\tilde{\mathbf{M}}\otimes\tilde{\mathbf{D}}\end{pmatrix}-H_0\begin{pmatrix}\tilde{\mathbf{E}}\\ \tilde{\mathbf{O}}\\ R\tilde{\mathbf{M}}\otimes\tilde{\mathbf{E}}\end{pmatrix} \tag{5}
$$

where

$$
\tilde{\mathbf{u}}=\{u_2,u_4,\cdots,u_{2N}\}^T,\ \tilde{\mathbf{p}}=\{p_1,p_3,\cdots,p_{2N-1}\}^T,\ \tilde{\mathbf{y}}=\{y_{1,2},y_{1,4},\cdots,y_{k,2N},y_{2,2},\cdots,y_{k,2N}\}^T,
$$

$$
\tilde{\mathbf{N}}=\{n_1,n_2,\cdots,n_k\}^T,\ \tilde{\mathbf{M}}=\{m_1,m_2,\cdots,m_k\}^T,\ \tilde{\boldsymbol{\Xi}}=\{1,1,\cdots,1\}^T
$$

$$
\mathbf{A}=\begin{bmatrix}10 & -12 & 2 & & & & \\ -1 & 11 & -11 & 1 & & & \\ & \ddots & \ddots & \ddots & \ddots & & \\ & & -1 & 11 & -11 & 1 & \\ & & & -1 & 11 & -11 & \\ & & & & -2 & 12 & \end{bmatrix}\quad
\mathbf{B}=\begin{bmatrix}-12 & 2 & & & & & \\ 11 & -11 & 1 & & & & \\ -1 & 11 & -11 & 1 & & & \\ & \ddots & \ddots & \ddots & \ddots & & \\ & & -1 & 11 & -11 & 1 & \\ & & & -2 & 12 & -10 & \end{bmatrix}
$$

$$
\tilde{\mathbf{C}}=\{10,-1,0,\cdots,0,0,0\}^T\quad \tilde{\mathbf{D}}=\{0,0,0,\cdots,0,1,-10\}^T\quad \tilde{\mathbf{E}}=\{1,1,1,\cdots,1,1,1\}^T\quad \mathbf{F}=\mathrm{diag}\left[n_1\mathbf{I},n_2\mathbf{I},\cdots,n_k\mathbf{I}\right].
$$

The symbol $\otimes$ defines the Kronecker product of a vector $\tilde{\mathbf{Z}} = \{z_1, z_2, \cdots, z_k\}^T$ and a matrix $\mathbf{G}$, that is, $\tilde{\mathbf{Z}} \otimes \mathbf{G} \equiv \{z_1\mathbf{G}, z_2\mathbf{G}, \cdots, z_k\mathbf{G}\}^T$.

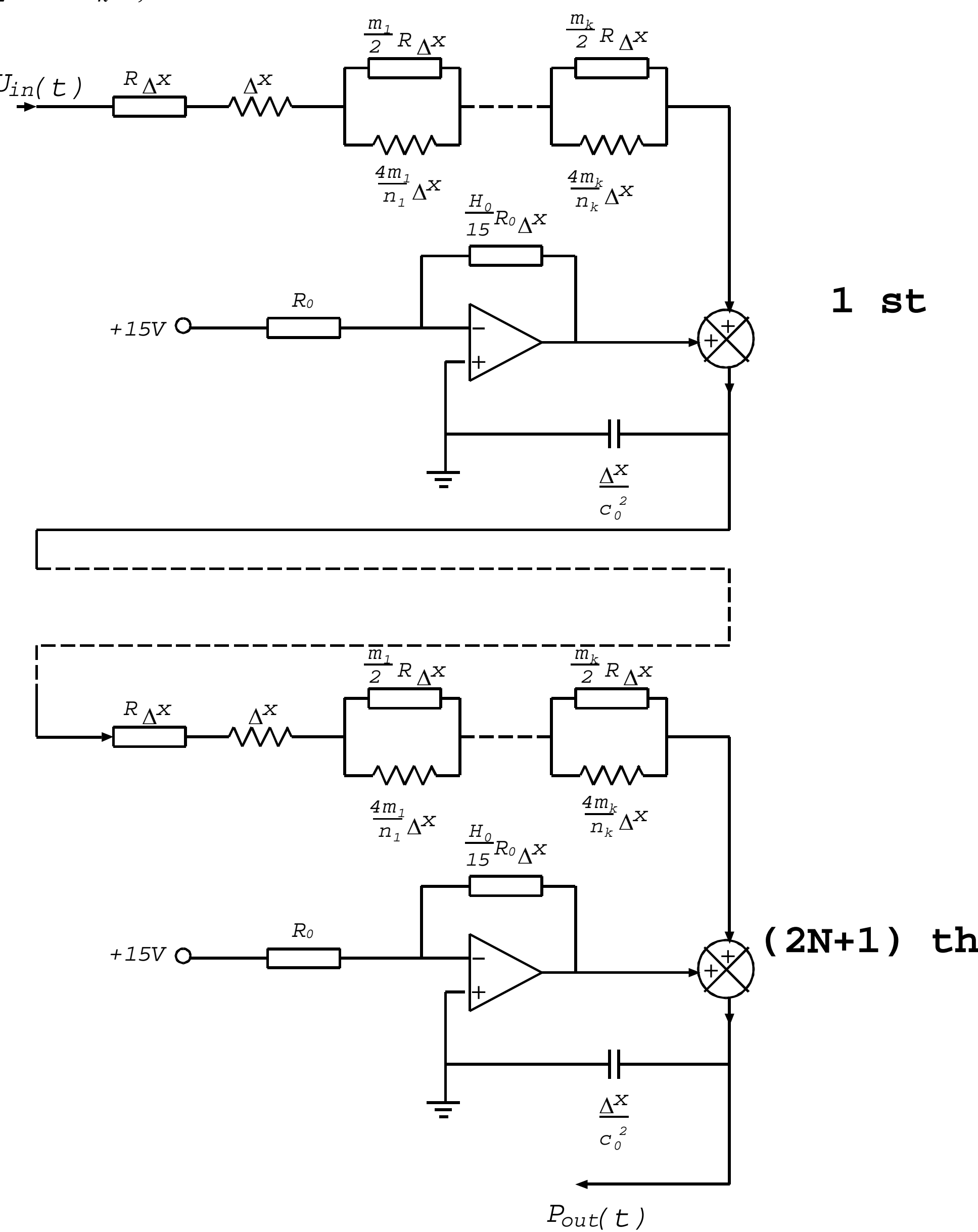

Figure 2: Equivalent electronic analogue circuit

## ELECTRONIC ANALOGUE

The new ordinary equation allows us to simulate the fluid transients by constructing an equivalent electronic analogue circuit as shown in Figure 2. The value and type for each electronic component are indicated.

For a section of transmission line of length $\Delta x$, this can be considered to offer a resistance due to the steady fluid friction, an inductor due to the fluid inertia and a capacitor due to the fluid compressibility. The frequency-dependent friction can simply be formed by a set of subcircuits, each of which is made of a resistor and an inductor hooked in parallel. The inverting operational amplifier indicates a constant pressure drop due to gravity for an inclined tube. On reading the analogue circuit, we may get the impression that, without the frequency-dependent friction, the circuit is almost identical to a $\pi$-type LC

Butterworth passive low-pass filter (Zverev, 1967; Horowitz and Hill, 1980). Over time, the waves produced by the changes of boundary conditions become smoother due to the suppression of high-frequencies superimposed on the waves. The parallel RL subcircuit is representative of frequency-dependent friction, which enhances the function of the low-pass filter by increasing the passing impedance of high-frequency signals.

The great advantage in using the analogue circuit to analyse the fluid transients has been found in the following three major points:

a) Quantitative analysis of analogue circuits can help us fully understand fluid transients using existing well-established electronic theory.
b) An equivalent electronic analogue circuit is able to provide a monitoring base, such as differentiating cavitation noise from system noise.
c) The low-pass filter design theory is able to modify the circuit to get more precise mathematical models, such as the frequency-dependent friction model.

## NUMERICAL RESULTS

The transient response of the finite element formulation was compared with that of the method of characteristics to demonstrate the accuracy of the Galerkin method based on previously published data (Lee et al. 1985), which are listed in Table 2 and shown a good agreement between experimental results and numerical computations by means of the method of characteristics for transient flow under negative pressure, in which the liquid can withstand vaporized tensile strengths lower than the vapor pressure. So, the results produced by the method of characteristics can be treated as an accurately analytical reference.

Table 2: Parameter list

| Time tolerance $(s)$ | Upstream pressure $(MPa)$ | Downstream pressure $(MPa)$ | Radius $r_0$ $(mm)$ | Length $L$ $(km)$ | Density $\rho$ $(kg/m^3)$ | Acoustic velocity $c_0$ $(m/s)$ | Viscosity $\mu$ $(cP)$ |
|---|---|---|---|---|---|---|---|
| 0.2 | 3 | 2 | 4 | 0.02 | 871 | 1392 | 50.518 |

It is to be noted at this stage that ordinary partial equations are based on the Galerkin finite element method in space domain only. If the finite element is to be adopted for the time domain also, the problem becomes unmanageable even on large computers. A compromise is to have another numerical scheme in the time domain. Although a formidable number of methods can be found for solving the ordinary differential equations (Hairer et al. 1987), the Dormand and Prince 5 th order integrator (1980) has been adopted here.

The comparison for $p_0$ is usually meaningful, but the value of $p_0$ is exclusive from this kind of arrangement. The interpolation using the Galerkin method

$$\int_{x_0}^{x_1} w_0^+ L_1(U, P, Y_i)dx = 0 \quad \text{gives} \quad \frac{dp_0}{dt} = -\frac{1}{2}\frac{dp_1}{dt} + \frac{3c_0^2}{4\Delta x}\left(U_{\text{in}} - u_2\right).$$

The result for $p_0$ is plotted and compared as shown in Figure 3.

## CONCLUSIONS

A finite element model with frequency-dependent friction has been developed. As has been displayed the finite element model retains the accuracy of the method of characteristics. Based on the finite element model, an electronic analogue methodology has been presented for the problem of fluid transients. The use of the equivalent electronic analogue circuit provides an alternative way to understand the mechanism of transient fluid flow and improve the frequency-dependent friction model mathematically.

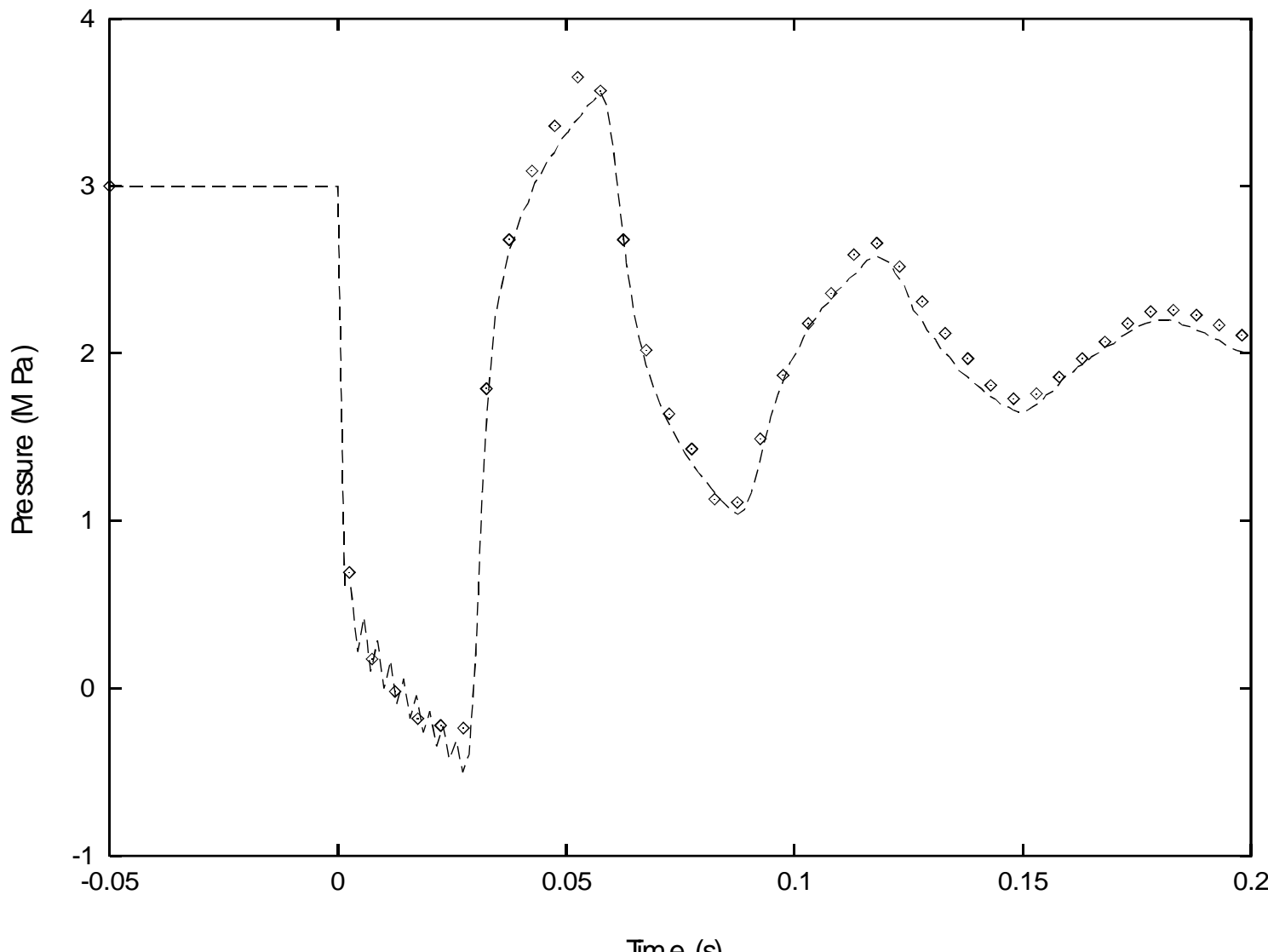

Figure 3: Finite element method (point) vs. method of characteristics (line)